# A Spin Glass State in $Ba_3TiRu_2O_9$

Loi T. Nguyen and R.J. Cava

Department of Chemistry, Princeton University, Princeton, New Jersey 08544, USA

**Abstract**

The magnetic properties of $Ba_3TiRu_2O_9$, whose crystal structure is based on stacked triangular planar lattices of $MO_6$ dimers and single $MO_6$ octahedra, are reported. The system is magnetically disturbed by a substantial amount of Ti/Ru chemical disorder. The Weiss temperature and effective magnetic moment were found to be -29.5 K and 1.82 $\mu_B$/f.u. respectively, and a bifurcation in the zero field cooled and field cooled magnetic susceptibility is observed below 4.7 K, suggesting that this is a compositionally-disordered spin-glass system. The material is a semiconductor with an activation energy for charge transport of approximately 0.14 eV.

**Keywords**: magnetic frustration, spin glass, ruthenium dimers, triangular lattice.

## 1. Introduction

Spin glass systems, with a disordered low temperature magnetic state often caused by atomic disorder[1],[2], have been of interest for many years[3]. Local small spin clusters are gradually formed when a spin-glass material is cooled, and, at the freezing temperature, the system is stuck in one metastable configuration of many degenerate ground states[4]. Ruthenium oxides can often display unusual magnetic and electrical properties due to the multiple degrees of freedom of the Ru 4$d$ electrons. $Sr_2RuO_4$, for example, becomes a superconductor below 1 K while $SrRuO_3$ is ferromagnetic at 160 K[5],[6], with strong spin orbit coupling is observed in $Ba_3CoRu_2O_9$[7] and magnetoelastic coupling in $Ba_3BiRu_2O_9$[8]. Systems that include Ru-Ru dimers can also exhibit interesting properties - $Ba_3CoRu_2O_9$ for example is antiferromagnetic with Weiss temperature of 93 K, and has robust coupling among orbital, spin and charge degrees of freedom[7] while $Ba_3MRu_2O_9$ (M=$La^{3+}$, $Nd^{3+}$ and $Y^{3+}$)[9] materials display ferromagnetic interactions within the Ru-Ru dimers and antiferromagnetic interactions between dimers. Here we report that the related material $Ba_3TiRu_2O_9$, isostructural with $Ba_3BiRu_2O_9$[8], based on $Ru_2O_9$ dimers in triangular planes, shows magnetic behavior that is dominated by strong chemical disorder. The bifurcation between its ZFC and FC susceptibility at 4.7 K indicates the presence of a low temperature spin-glass state with a freezing temperature $T_f$ = 4.7 K.

## 2. Experimental

A polycrystalline sample of $Ba_3TiRu_2O_9$ was synthesized by solid-state reaction using $BaCO_3$, $RuO_2$, and $TiO_2$ (Alfa Aesar, 99.9%). The mixture of reagents in stoichiometric ratios was heated in air at 1100 °C for 12 hours, reheated at 1100 °C and then at 1200 °C for 12 hours. The phase purity and crystal structure of our sample of $Ba_3TiRu_2O_9$ was determined using powder X-ray diffraction (PXRD) using a Bruker D8 Advance Eco with Cu Kα radiation and a LynxEye-XE detector. The structural refinements were performed with *GSAS*.[10] Crystal structure drawings were created by using the program *VESTA*.[11]

The magnetic susceptibility of $Ba_3TiRu_2O_9$ was measured by a Quantum Design Physical Property Measurement System (PPMS) DynaCool equipped with a VSM option. For those measurements, the sample was ground into powder and placed in plastic capsules. The magnetic susceptibility, defined as M/H, where M is the sample magnetization and H is the applied field, was measured at the field of H = 1 kOe from 1.7 K to 300 K. Zero-field cooled (ZFC) and field-cooled (FC) magnetization data were measured from 1.7 K to 20 K at an applied field of 30 Oe for $Ba_3TiRu_2O_9$. The resistivity of $Ba_3TiRu_2O_9$ was measured by the DC four-contact method in the temperature range 250 K to 330 K with the PPMS. The sample was pressed, sintered and cut into pieces with the approximate sizes $1.0 \times 2.5 \times 1.5$ mm$^3$. Four Pt contact wires were connected to the samples using silver paint.

## 3. Results

The powder X-ray diffraction pattern and structural refinement results for $Ba_3TiRu_2O_9$ are shown in **Figure 1a**. The crystal structure of $Ba_3TiRu_2O_9$ is shown in **Figure 1b**. The lattice parameters and atomic positions reported in **Table 1** are similar to those reported previously[12]. $Ba_3TiRu_2O_9$ adopts a hexagonal structure with the space group $P6_3/mmc$ (No.194). Ti occupies 51.2% of the isolated octahedra, while in the dimers there is an occupancy of 17.7% Ti. This indicates that although Ti prefers the isolated octahedra, while Ru prefers the face-shared octahedra, there is a substantial amount of chemical disorder present. We attribute this disorder to the similarities in the charge and radii of the Ti and Ru ions.[13]

The temperature-dependent magnetic susceptibility of $Ba_3TiRu_2O_9$ and its reciprocal are plotted in **Figure 2**. The Curie-Weiss fitting for data from 75-300 K yields a Weiss temperature of -29.5

K, effective magnetic moment of 1.82 $\mu_B$/f.u., and a temperature-independent paramagnetic susceptibility of $9.04\times10^{-4}$ emu $Oe^{-1}$ f.u.$^{-1}$. The observed effective moment is 1.82 $\mu_B$/f.u, which is smaller than what is typically expected for S = 1 $Ru^{4+}$ ions [15],[16],[17]. This may be due to the effect of spin orbit coupling, which can partially split the three degenerate $t_{2g}$ states into two states with the same energy and a third one at higher energy, as is commonly seen in ruthenates and iridates[18],[19],[20]. Another possibility is that only Ru ions either in the isolated octahedra or in the dimers are magnetic due to different coordination environments.

The inset shows the bifurcation of FC and ZFC magnetic susceptibility at the temperature 4.7 K in an applied DC field of 30 Oe. Thee data imply that the material is a spin glass with the freezing temperature 4.7 K. Similar spin glass behavior has been seen in $Ba_3FeRu_2O_9$[14], with the disorder between Fe and Ru leading to a spin glass state with a freezing temperature of 25 K. ZFC/FC DC magnetic susceptibilities were measured at different fields to investigate the spin glass behavior in $Ba_3TiRu_2O_9$. As shown in **Figure 3a**, freezing temperature is determined to be 4.7 K at the applied field of 60 Oe. When the field is increased to 2.5 kOe, $T_f$ is shifted down to 3.8 K. At the field of 1 T, it is moved to 2.5 K. Above 2 T, the freezing temperature is further suppressed below 2 K and is not detected in our measurements. This is similar to the cluster spin glass behavior seen in $Cr_{0.5}Fe_{0.5}Ga$[21]. The resistivity of $Ba_3TiRu_2O_9$ as a function of reciprocal temperature is plotted in **Figure 3b**. $Ba_3TiRu_2O_9$ is nonmetallic, similar to what is observed for $Ba_3BiRu_2O_9$.[8] The activation energy is calculated to be $E_a$ = 0.14 eV in the temperature range from 250-300 K.

## 4. Conclusions

$Ba_3TiRu_2O_9$ crystallizes in a 6-layer hexagonal unit cell in the *P6$_3$/mmc* space group. The material has high resistivity, indicating that it is semiconducting; the measured activation energy for charge transport is approximately 0.14 eV. The bifurcation in the zero field cooled (ZFC) and field cooled (FC) magnetic susceptibilities indicates the presence of a spin glass transition at about 4.7 K. This may come from the fact that material displays substantial structural disorder, although there may also be some contribution from the magnetically frustrating triangular arrangement of the Ru-Ru dimers and the single $RuO_6$ octahedra. The lower than expected spin on the $Ru^{4+}$ present suggests that theoretical modelling should best be performed to understand the magnetism of this material.

## 5. Acknowledgements

This work was supported by the Basic Energy Sciences of the Department of Energy grant number DE-FG02-08ER46544 through the Institute for Quantum Matter at Johns Hopkins University.

**Table 1. Atomic coordinates and equivalent isotropic displacement parameters of $Ba_3TiRu_2O_9$, space group $P6_3/mmc$ ( No. 194).**

| Atom | Wyckoff. | Occ. | x | y | z | $U_{iso}$ |
|------|----------|------|---|---|---|-----------|
| Ba1 | 2b | 1 | 0 | 0 | ¼ | 0.0015(5) |
| Ba2 | 4f | 1 | ⅓ | ⅔ | 0.09153(8) | 0.0023(3) |
| Ti1 | 2a | 0.512(8) | 0 | 0 | 0 | 0.0076(5) |
| Ru2 | 2a | 0.488(8) | 0 | 0 | 0 | 0.0076(5) |
| Ti2 | 4f | 0.177(6) | ⅓ | ⅔ | 0.8419(1) | 0.0014(5) |
| Ru1 | 4f | 0.823(6) | ⅓ | ⅔ | 0.8419(1) | 0.0014(5) |
| O1 | 6h | 1 | 0.5105(8) | 0.021(2) | ¼ | 0.0059(3) |
| O2 | 12k | 1 | 0.8369(7) | 0.674(1) | 0.0801(4) | 0.0060(2) |
| a = 5.71392(4) Å, c = 14.0351(1) Å ||||||| 
| $R_{wp}$ = 12.03%, $R_p$ = 8.76%, $R_F^2$ = 5.05% |||||||

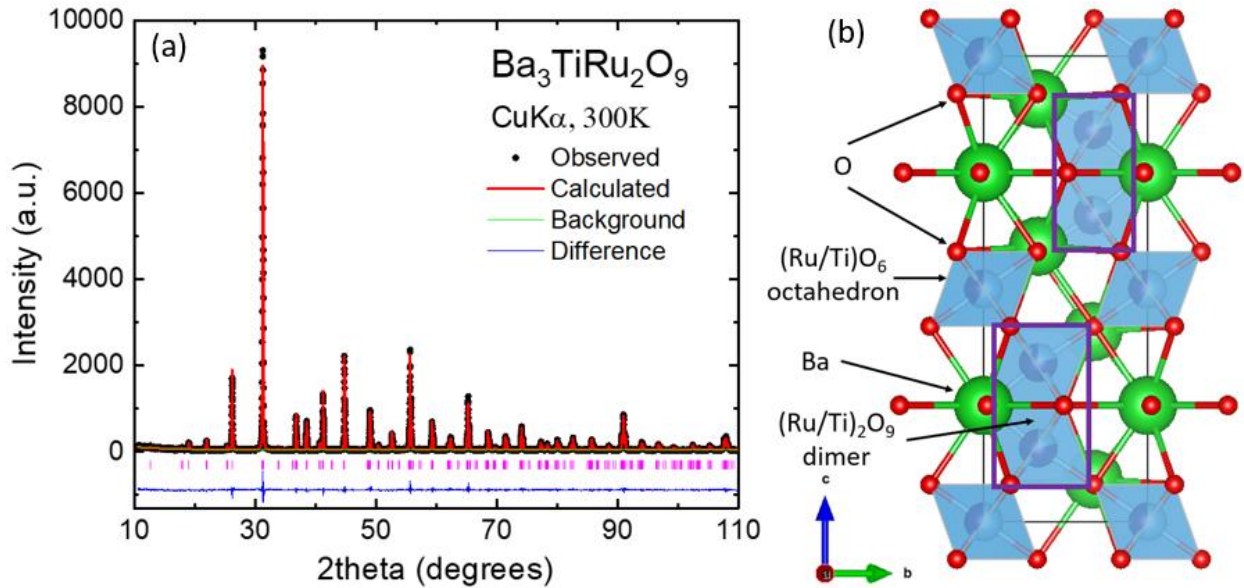

**Figure 1:** (a) Rietveld powder x-ray diffraction refinement of $Ba_3TiRu_2O_9$ in space group *P6_3/mmc*. The observed X-ray pattern is shown in black, calculated in red, difference ($I_{obs}$-$I_{calc}$) in blue, the background in green, and tick marks denote allowed peak positions in pink. $R_p = 0.0876$, $R_{wp} = 0.1203$, $\chi^2 = 1.480$. (b) Crystal structure of $Ba_3TiRu_2O_9$ in space group *P6_3/mmc*. The isolated $(Ru/Ti)O_6$ octahedra (blue) are corner-shared to the $(Ru/Ti)_2O_9$ dimers (blue with purple rectangles) to form what is frequently referred to as a "6-layer" hexagonal perovskite structure. Barium is labeled in green and oxygen is in red.
9

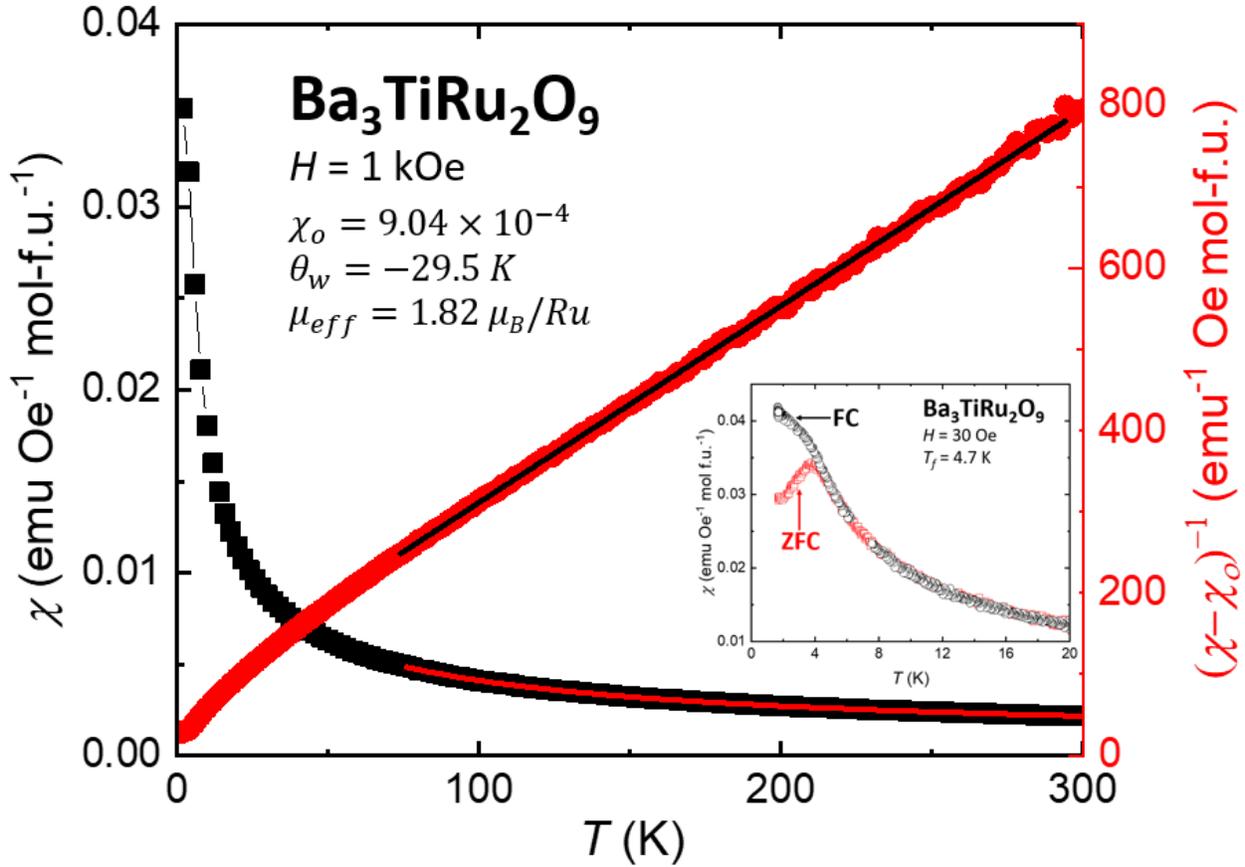

**Figure 2: The temperature dependence of the magnetic susceptibility (black squares) and the inverse of the difference between the magnetic susceptibility and the temperature independent magnetic susceptibility (red circles) for $Ba_3TiRu_2O_9$. The applied field is 1 kOe. The red solid line is the susceptibility fit calculated from Curie-Weiss law from 75-300 K. The inset shows the bifurcation of the field cooled (FC) and zero field cooled (ZFC) DC susceptibility in an applied field of 30 Oe. The spin glass transition temperature is determined to be 4.7 K.**

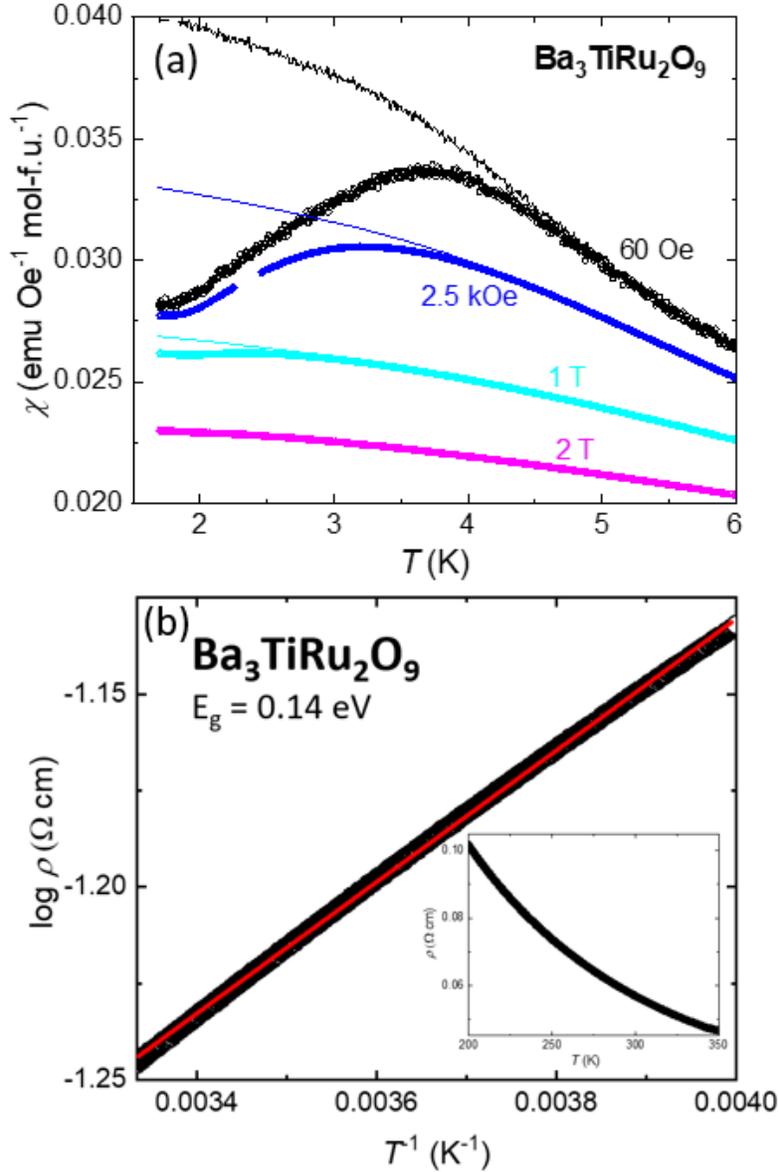

**Figure 3:** (a) The temperature dependence of the ZFC and FC magnetic susceptibilities measured under different applied fields for Ba$_3$TiRu$_2$O$_9$. The freezing temperatures are decreasing as the fields increase. (b) Temperature-dependent resistivity of a sintered pellet of Ba$_3$TiRu$_2$O$_9$ as a function of inverse temperature, in log-form. Data in the elevated temperature regime (250-300 K) was fit to the model $\rho = \rho_o e^{\frac{E_a}{k_b T}}$ (red line) with E$_a$ = 0.14 eV. Inset – The raw resistivity data from 200-350 K.